\documentclass[aps,prl,twocolumn,showpacs,citeautoscript,longbibliography]{revtex4-1}

\usepackage{graphicx}  
\usepackage{dcolumn}   
\usepackage{bm}        
\usepackage{amssymb}   
\usepackage[none]{hyphenat}  
\usepackage{nicefrac}  
\usepackage{amsmath}   
\usepackage{footmisc}  
\usepackage{latexsym}  
\usepackage{romannum}  
\usepackage[colorlinks,linkcolor=blue,anchorcolor=blue,citecolor=blue,urlcolor=blue]{hyperref}

\newcommand{\xv}{\mathbf{x}}
\newcommand{\ev}{\mathbf{\hat{e}}}
\renewcommand{\Im}{\mathrm{Im}}
\renewcommand{\Re}{\mathrm{Re}}
\newcommand{\eps}{\varepsilon}

\begin{document}

\title{Circularly polarized thermal radiation from nonequilibrium coupled antennas}
\author{Chinmay Khandekar}
\email{ckhandek@purdue.edu}
\affiliation{Birck Nanotechnology Center, School of Electrical and Computer Engineering, College of Engineering, Purdue University, West Lafayette, Indiana 47907, USA}

\author{Zubin Jacob}
\email{zjacob@purdue.edu}
\affiliation{Birck Nanotechnology Center, School of Electrical and Computer Engineering, College of Engineering, Purdue University, West Lafayette, Indiana 47907, USA}

\date{\today}

\begin{abstract}
  Circularly polarized light can be obtained by using either
  polarization conversion or structural chirality. Here we reveal a
  fundamentally unrelated mechanism of generating circularly polarized
  light using coupled nonequilibrium sources. We show that thermal
  emission from a compact dimer of subwavelength, anisotropic antennas
  can be highly circularly polarized when the antennas are at unequal
  temperatures. Furthermore, the handedness of emitted light is
  flipped upon interchanging the temperatures of the antennas, thereby
  enabling reconfigurability of the polarization state lacked by most
  circularly polarized light sources. We describe the fundamental
  origin of this mechanism using rigorous fluctuational electrodynamic
  analysis and further provide practical examples for its experimental
  implementation. Apart from the technology applications in
  reconfigurable devices, communication, and sensing, this work
  motivates new inquiries of angular momentum related thermal
  radiation phenomena using thermal nonequilibrium, without applying
  magnetic field.
\end{abstract}

\maketitle

\section{Introduction}

Circularly polarized (CP) light has recently acquired great attention
in context of spin-controlled
nanophotonics~\cite{le2015nanophotonic,mitsch2014quantum},
spintronics~\cite{vzutic2004spintronics} and chiral quantum
optics~\cite{lodahl2017chiral} where its spin angular momentum is
harnessed for engineering spin-dependent light matter interactions at
nanoscale. Given its fundamental and technological importance, there
is a strong demand for CP light sources having high purity and
compactness with/without reconfigurability of the polarization state.

One approach to obtain CP light is by passing unpolarized light
through a linear polarizer followed by an optimized polarization
conversion device such as a
metasurface~\cite{zhu2013linear,pfeiffer2014high,jiang2017ultra},
which can preferentially convert it into right circularly polarized
(RCP) light or left circularly polarized (LCP) light. Another more
fundamental approach utilizes structural (geometric/material)
chirality at the level of the source, examples of which include a long
list of
electroluminescent~\cite{nishizawa2017pure,asshoff2011spintronic,
  di2017high,zhao2016circularly,zhang2014electrically} and
photoluminescent~\cite{kumar2015circularly,sanchez2015circularly,
  konishi2011circularly,maksimov2014circularly} CP light sources.
Apart from these approaches, CP light generation via thermal radiation
(incandescence) has been demonstrated experimentally using the same
underlying mechanisms~\cite{wadsworth2011broadband,
  shitrit2013spin}. These are few representative examples of all types
of CP-light sources which are fundamentally based on either
polarization conversion or structural chirality.

In this work, we demonstrate a different mechanism of generating CP
light based on near-field coupling between nonequilibrium sources. It
is unrelated to polarization conversion or chirality and cannot be
described using those concepts. This mechanism is useful because
practically, it solves the challenging problem of reconfigurability of
the polarization state lacked by most CP light sources. Fundamentally,
it reveals an unforeseen connection between thermal nonequilibrium and
angular momentum of light.

\begin{figure}[t!]
  \centering \includegraphics[width=\linewidth]{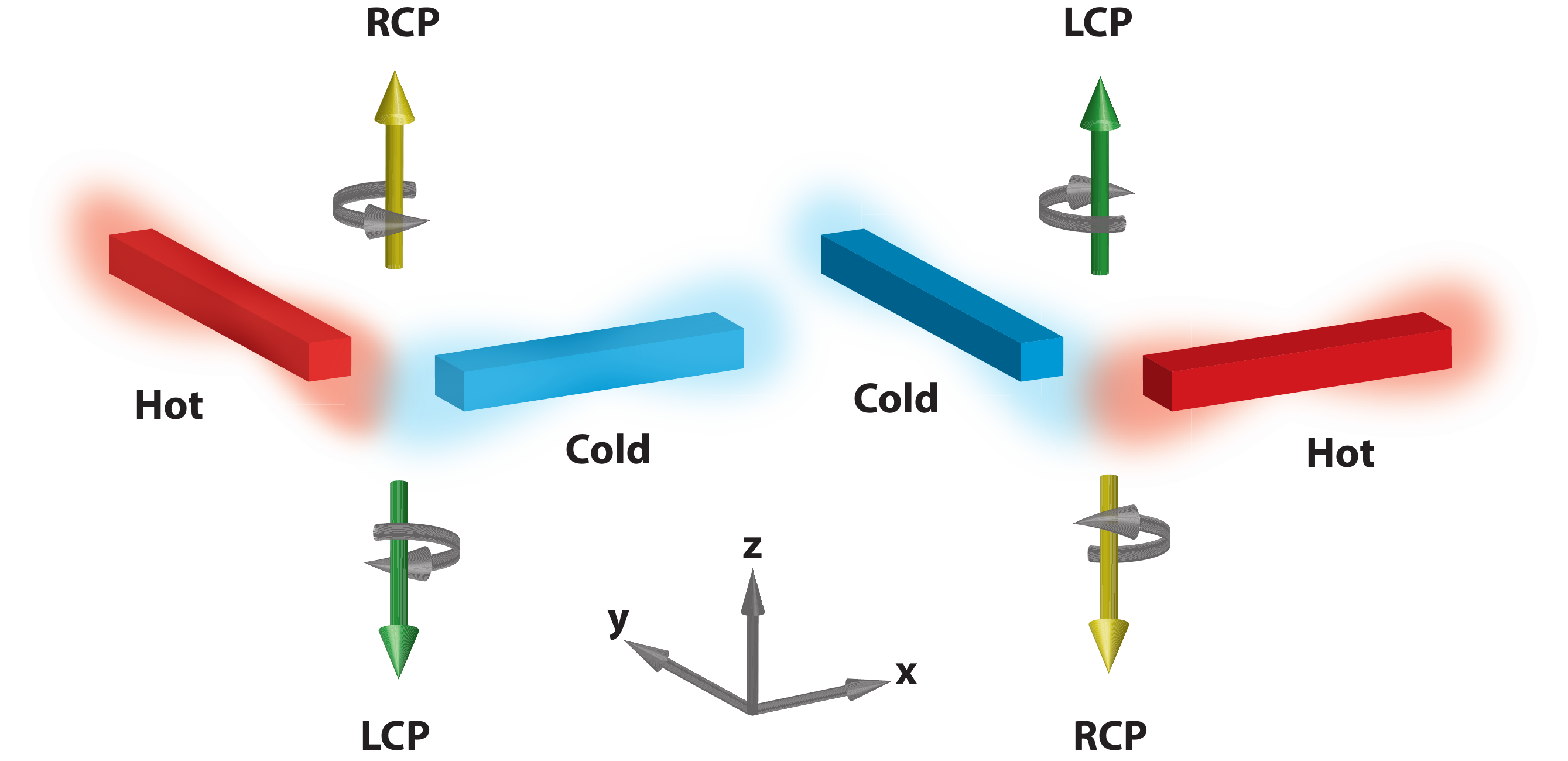}
  \caption{A pair of anisotropic shaped, subwavelength (dipolar),
    coupled antennas at unequal temperatures can emit strongly
    circularly polarized light in suitable directions (upper and lower
    hemispheres). Handedness/directionality of emitted polarized
    radiation is switched upon interchanging the temperatures of two
    antennas.}
  \label{schematic}
\end{figure}

We consider a dimer system of two perpendicularly oriented,
subwavelength, anisotropic shaped antennas depicted in
figure~\ref{schematic}. We analyze the spin angular momentum of
thermal emission from the dimer using fluctuational
electrodynamics. We note that this approach~\cite{ott2018circular}
remains largely unexplored in the field of thermal radiation despite
many decades of separate works on radiative heat
transfer~\cite{rytov1959theory,landau1980statistical} and angular
momentum of non-thermal
light~\cite{barnett2016optical,allen1992orbital}. The fluctuational
electrodynamic analysis reveals that suitable correlations (imaginary
valued) between orthogonal components of thermally fluctuating sources
are necessary for emission of CP light. We find that the strong
correlations are realized through near field interactions between the
antennas when they are held at unequal temperatures
(nonequilibrium). On the other hand, antennas at equal temperatures
emit very weakly polarized thermal radiation. Interestingly, for
nonequilibrium antennas, the handedness of emitted radiation is
flipped upon interchanging the temperatures of the antennas thus
enabling reconfigurability of the polarization state lacked by most CP
light sources~\cite{shitrit2013spin,wadsworth2011broadband}. We
further explore the experimental feasibility of observing this
phenomenon with dimers of antennas made of Silicon Carbide (SiC) and
Indium Phosphide (InP), amongst many other plasmonic/polaritonic
materials. The resonantly enhanced near-field interaction between the
localized dipolar modes of the antennas, leads to reasonably strong
mid-infrared CP light emission under temperature difference of $\Delta
T\sim 30$K across separation of $d \sim 1\mu$m between the
antennas. This should be experimentally realizable in the near future
since a significant experimental progress in measuring near-field
radiative heat transfer has facilitated exploration of large
temperature differences at nanoscale and has verified the validity of
fluctuational electrodynamic
theory~\cite{song2015near,kim2015radiative,tervo2018near}.  We further
provide another conceptual example system based on plasmonic
nanolaser, operated well below lasing threshold, where amplified
spontaneous emission is described using fluctuational
electrodynamics. That example demonstrates that the proposed mechanism
is not limited to `thermal' nonequilibrium but can also be implemented
in other forms such as population inversion of gain atoms, paving the
way for CP light emitting LEDs.


In comparison to other solid-state designs that can emit CP thermal
radiation~\cite{wadsworth2011broadband,shitrit2013spin,wu2014spectrally,
  yin2013interpreting}, the proposed mechanism offers a practical
advantage of temperature-based polarization reconfigurability. The
underlying fluctuational electrodynamic analysis sheds additional
light on previous designs and expands the design space by
incorporating nonequilibrium systems. In the field of thermal
radiation, the proposed mechanism reveals a non-intuitive fundamental
connection between thermal nonequilibrium and angular momentum of
light. While almost all studies of angular momentum of light are
primarily limited to non-thermal
light~\cite{bliokh2017optical,bliokh2013dual, van2016universal}, few
recent studies~\cite{moncada2015magnetic,ott2018circular} have started
to explore related thermal radiation phenomena. But they require
application of magnetic field. Our work suggests the possibility of
exploring such phenomena using thermal nonequilibrium without applying
magnetic field. In context of using nonequilibrium systems for shaping
thermal radiation, other recent works have explored directional
emissivity~\cite{jin2016temperature,han2010control}, nontrivial
thermal forces and torques~\cite{reid2017photon} and enhanced
emissivity from nonequilibrium antennas~\cite{sakat2018enhancing}. Our
work conveys that one can take advantage of temperature-based
reconfigurability in nonequilibrium systems, for numerous applications
in communications, sensing and detection technologies.


We divide the paper into following parts. In Sec.~\ref{sec:dimer},
we demonstrate circular polarization (spin angular momentum) of
thermal emission from the dimer system of Fig.\ref{schematic} using
fluctuational electrodynamics. We describe the fundamental origin of
CP light emission from nonequilibrium antennas and further provide
general design guidelines for systems more complicated than the
proposed dimer. In Sec.~\ref{sec:example}, we explore the practical
implementation of the proposed mechanism by considering realistic SiC
and InP antenna dimers and a conceptual plasmonic nanolaser system as
another viable option.  Finally, we conclude by highlighting the
fundamental and technological relevance of this work and future
research directions in Sec.~\ref{sec:conclusion}.

\section{Dimer of antennas}
\label{sec:dimer}

\emph{Theory:} Thermal radiation consists of electromagnetic fields
generated by thermal, stochastic motion of charges. These can be
calculated from Maxwell's equations by introducing fluctuating
currents provided they satisfy specific fluctuation-dissipation
relations~\cite{rytov1959theory,novotny2012principles}. To study
thermal emission from a dimer of subwavelength antennas, we consider
one variant of this fluctuational electrodynamic approach termed as
finite dipole model where thermal sources are randomly fluctuating
dipole moments of dipolar sources, satisfying specific
fluctuation-dissipation
relations~\cite{ben2011many,domingues2005heat,joulain2005surface}. We
consider the geometry depicted in Fig.\ref{schematic} consisting of
two subwavelength (dipolar), anisotropic antennas placed near the
origin in vacuum with their long axes oriented along arbitrary $\ev_1$
and $\ev_2$ directions. The antennas have temperatures $T_1, T_2$ and
vacuum polarizabilities $\alpha_1, \alpha_2$ along their orientations
respectively. The fluctuating dipole moments of these dipolar antennas
in the absence of any interactions are given by $p_1 \ev_1$ and $p_2
\ev_2$ which satisfy the following fluctuation dissipation relations
in the frequency domain~\cite{ben2011many,domingues2005heat,
  joulain2005surface}:
\begin{align}
  \langle p_j^*(\omega)p_k(\omega')\rangle =
  \frac{2\eps_0}{\omega}
  \text{Im}\{\alpha_j \} \Theta(\omega,T_j) \delta(\omega-\omega')\delta_{jk}
\label{pfdt}  
\end{align}
Here $\{j,k\}=\{1,2\}$, $\Theta(\omega,T)=\hbar\omega/2+
\hbar\omega/[\mathrm{exp}(\hbar\omega/k_{B}T)-1]$ is the mean thermal
energy of a Harmonic oscillator of frequency $\omega$. $T_j$ is the
temperature of dipolar object and $\langle ... \rangle$ denotes
statistical ensemble average. When the dipoles are placed close to
each other, the interactions between them lead to effective dipole
moments:
\begin{align}  
\tilde{p}_j = p_j + \eps_0\alpha_j [\mathbf{E}_{jk}(\tilde{p}_k)\cdot
  \ev_j]
\label{pf}
\end{align}
for $j\neq k$. $\mathbf{E}_{jk}(\tilde{p}_k)$ denotes the electric
field at dipole $j$ due to dipole $k$. It is straightforward to
calculate the far field thermal radiation as well as the near-field
induced correlations of the effective dipole moments using the general
expressions,
\begin{align}
\label{Efield}
\mathbf{E}(\omega) &= \frac{k_0^3}{4\pi\eps_0}\frac{e^{ik_0 R}}{(k_0
  R)} \bigg[ [\ev_j-(\ev_j\cdot\ev_R)\ev_R]\nonumber \\ &+ \frac{ik_0
    R -1}{(k_0 R)^2}[\ev_j-3(\ev_j\cdot\ev_R)\ev_R] \bigg] \tilde{p}_j
\\ \mathbf{H}(\omega) &= \frac{k_0^3}{4\pi\eps_0}\frac{e^{ik_0
    R}}{(k_0 R)} \sqrt{\frac{\eps_0}{\mu_0}} \bigg[1 + \frac{i}{k_0 R}
  \bigg] (\ev_R \times \ev_j) \tilde{p}_j
  \label{Hfield}
\end{align}
for the electromagnetic fields at a point $R\ev_R$ due to a single
dipole $\tilde{p}_j \ev_j$ at the origin with $k_0=\omega/c$ being the
vacuum wavevector.

The intensity flux of thermal radiation is given by the Poynting
vector in the far field ($\Re\{\mathbf{E}^*(\omega) \times
\mathbf{H}(\omega)\}$) and the degree of circular polarization is
measured by the spin-angular-momentum density of thermal emission. By
generalizing the definitions employed for non-thermal light in several
works~\cite{barnett2016optical,berry1998paraxial,bliokh2013dual} to
thermal radiation~\cite{ott2018circular}, we write the spectral energy
density $\langle W(\omega) \rangle$ and spin angular momentum density
$\langle \mathbf{S}(\omega) \rangle$ of emitted light in vacuum as:
\begin{align}
  &\langle W(\omega) \rangle = \frac{1}{2} \langle \eps_0
  \mathbf{E}^*(\omega) \cdot \mathbf{E}(\omega) +
  \mu_0\mathbf{H}^*(\omega) \cdot \mathbf{H}(\omega) \rangle
  \\ &\langle \mathbf{S}(\omega) \rangle = \frac{1}{2\omega} \Im\{
  \langle \eps_0 \mathbf{E}^*(\omega) \times \mathbf{E}(\omega) +
  \mu_0 \mathbf{H}^*(\omega)\times \mathbf{H}(\omega)
  \rangle\} \label{spin}
\end{align}
Note the use of $\langle...\rangle$ which denotes the statistical
ensemble average of physical quantities in context of thermally
generated radiation. We further define a dimensionless vector quantity
called as spectral thermal spin,
\begin{align}
\mathbf{S}_T(\omega) = \frac{\omega \langle \mathbf{S}(\omega)
  \rangle}{\langle W(\omega) \rangle}
\label{Stdef}
\end{align}
whose magnitude in a given direction lies between $[-1,1]$ with $-1$
denoting pure LCP light and $+1$ denoting pure RCP light along that
direction. In an actual experiment, this direction can be along the
axis of a detector. Since the quantity given by Eq.\eqref{spin}
remains largely unexplored in the field of thermal radiation and may
not be familiar to researchers working on thermal radiation and heat
transfer topics, we point out its connection with closely related
well-known concepts. In particular, one can also understand the spin
of thermal radiation in more familiar language of Stokes polarimetry.
The Stokes $S_3$ parameter describes the circular polarization of the
fields lying transverse to a given propagation direction. If one
calculates the Stokes $S_3$ parameters for thermally generated fields
in all three orthogonal coordinate planes~\cite{setala2002degree}, one
retrieves a form similar to equation~\eqref{spin}. It is also
well-known that circularly polarized laser light can impart angular
momentum to small, absorptive particles in its path causing them to
rotate about their own axis~\cite{angelsky2012circular}. It is then
meaningful to calculate the spin angular momentum of light
[Eq.~\eqref{spin}] since optical torque on these particles is
proportionate to it~\cite{canaguier2013force,nieto2015optical}.

We calculate the spectral thermal spin given by Eq.~\eqref{Stdef} for
the example geometry of two dipolar antennas placed close to each
other near the origin with their centers at $\mathbf{x}_1$ and
$\mathbf{x}_2=\mathbf{x}_1+ d\ev_d$. The fluctuating dipole moments
and associated correlations such as $\langle
\tilde{p}_j^*(\omega)\tilde{p}_k(\omega) \rangle$ are calculated using
equations (\ref{pfdt}-\ref{Hfield}). The calculation of
electromagnetic fields and other related quantities in the far-field
at a point $R\ev_R$ is simplified since only the leading order terms
$\mathcal{O}(\frac{1}{k_0R})$ are important. The distances between
dipoles and $R\ev_R$ in the far-field are
$R_j=|R\ev_R-\mathbf{x}_j|=R-\mathbf{x}\cdot\ev_R$. Using simple
algebraic manipulations, we derive the following expression for the
spectral thermal spin in the far field at a point $R\ev_R$:
\begin{widetext}
\begin{gather}
\mathbf{S}_T(\omega) =
\frac{2\Im\{e^{ik_0(\mathbf{x}_1-\mathbf{x}_2)\cdot\ev_R}\langle
  \tilde{p}_1^*(\omega)\tilde{p}_2(\omega)\rangle\}
        [(\ev_1\times\ev_2)\cdot \ev_R]\ev_R}{\sum_{j,k=1,2}
  e^{ik_0(\mathbf{x}_j-\mathbf{x}_k)\cdot\ev_R}\langle
  \tilde{p}_j^*(\omega)\tilde{p}_k(\omega) \rangle [(\ev_j\cdot\ev_k)
    - (\ev_j\cdot\ev_R)(\ev_k\cdot\ev_R)]} \label{St} \\ \langle
\tilde{p}_1^*(\omega)\tilde{p}_2(\omega) \rangle =
\frac{2\eps_0}{\omega |D|^2}[(\alpha_2 k_0^3
  \kappa)\Im\{\alpha_1\}\Theta(\omega,T_1)+ (\alpha_1 k_0^3 \kappa)^*
  \Im\{\alpha_2\}\Theta(\omega, T_2)], \hspace{15pt} D =
1-\alpha_1\alpha_2 k_0^6\kappa^2\nonumber \\ \langle
|\tilde{p}_1(\omega)|^2\rangle=
\frac{2\eps_0}{\omega|D|^2}[\Im\{\alpha_1\} \Theta(\omega,T_1) +
  \Im\{\alpha_2\}\Theta(\omega,T_2)|\alpha_1 k_0^3 \kappa|^2],
\hspace{15pt} \langle |\tilde{p}_2(\omega)|^2\rangle = \langle
|\tilde{p}_1(\omega)|^2\rangle (1 \leftrightarrow 2), \nonumber
\end{gather}
\end{widetext}
The dimensionless near field coupling $\kappa$ between the two dipoles
is given by:
\begin{align*}
\kappa = &\frac{e^{ik_0 d}}{4\pi(k_0 d)^3} \bigg[ (k_0d)^2[(\ev_1\cdot
    \ev_2)-(\ev_1\cdot \ev_d)(\ev_2\cdot \ev_d)] \\ &+
  (ik_0d-1)[(\ev_1\cdot \ev_2)-3(\ev_1\cdot \ev_d)(\ev_2\cdot
    \ev_d)] \bigg]
\end{align*}
Equation (\ref{St}) is the central result of this work which offers
many analytic insights regarding generation of CP light as discussed
below.

{\bf Necessary condition for emission of CP light:} It follows from
Eq.\eqref{St} that the spin of far-field thermal emission is always
radial which we identify as radial thermal spin
$S_R=\mathbf{S}_T\cdot\ev_R $. Evidently $S_R = 0$ in all directions
($\ev_R$) when $\ev_1=\ev_2$ (parallel dipoles) and
$\Im\{e^{ik_0(\xv_1-\xv_2)\cdot\ev_R}
\langle\tilde{p}_1^*(\omega)\tilde{p}_2(\omega)\rangle\} = 0$ (no
correlations). The latter condition can be used to find the necessary
condition for emission of CP radiation from an arbitrary body.  We
note that if instead of two physically separate antennas, a single
dipolar object with dipole moments $p_j \ev_j$ for $j \in [x,y,z]$ is
considered, then it follows that the correlations $\Im\{
\langle\tilde{p}_j^*(\omega)\tilde{p}_k(\omega)\rangle\}$ between
these orthogonal components are necessary to produce CP radiation in
the far-field. Since it is not possible to have nonzero thermal spin
with zero correlations between orthogonal components of underlying
fluctuating sources, it follows that this is a necessary condition for
generation of CP light. This condition can be generalized to arbitrary
bodies by discretizing them into subvolumes much smaller than the
emission wavelength and conceptualized as electric point dipoles. This
approach is known as thermal discrete dipole
approximation~\cite{edalatpour2014thermal,edalatpour2015convergence}.
It follows that the correlations
$\Im\{\langle\tilde{p}_j^*(\omega)\tilde{p}_k(\omega)\rangle\}$
between the orthogonal components of the effective dipole moments of
these subvolumes are necessary to emit CP light.


This finding that the imaginary valued correlations are necessary for
emission of CP light is also an important result of this work. It is
insightful for designing CP thermal light sources. For instance, the
environment surrounding a dipolar thermal emitter can be engineered
such that the imaginary valued correlations between the effective
dipole moment components are nonzero and consequently, light emitted
is circularly polarized. We note that, for arbitrary bodies of
non-dipolar nature, the necessary condition may not be sufficient
since there can be cancellation of spin due to contribution from
dipoles of many tiny subvolumes, requiring a full calculation to infer
the circular polarization of total radiation emitted by such a
body. Nonetheless, such calculations could be performed with advanced
computational tools~\cite{reid2015efficient,edalatpour2015convergence}
and using definition~\eqref{spin} above. In this work, we first focus
on a simple dipolar dimer which is easier to understand, analyze and
optimize.

{\bf Optimum design for maximum purity CP light along normal
  direction:} We now find a design that emits maximum purity ($S_R=\pm
1$) CP thermal radiation. To simplify the optimization, we consider
the example geometry of Fig.\ref{schematic} with $\ev_1=\ev_x$,
$\ev_2=\ev_y$ (antennas lying in the $xy$-plane) and focus on thermal
emission in $\ev_z$ direction. Since the phase factor
$e^{ik_0(\xv_1-\xv_2)\cdot\ev_z}=1$ for $\ev_z$ direction, radial spin
$S_R \sim \Im\{
\langle\tilde{p}_1^*(\omega)\tilde{p}_2(\omega)\rangle\}$ and useful
analytical expressions can be obtained. Considering a practically
relevant situation of equal vacuum polarizabilities
$\alpha_1=\alpha_2=\alpha$, one finds that the radial thermal spin
$S_R \sim \Im[\alpha k_0^3 \kappa+(\alpha k_0^3 \kappa)^*
  \frac{\Theta(\omega,T_2)}{\Theta(\omega,T_1)}]$. This yields $S_R=0$
when both the dipolar antennas are at equal temperatures
$T_1=T_2$. Therefore, we consider nonequilibrium configuration ($T_1
\neq T_2$) for which the radial thermal spin in the (normal) direction
$\ev_R=\ev_z$ is:
\begin{align*}
S_R = \frac{2\Im\{\tilde{\alpha}+
  \tilde{\alpha}^*\frac{\Theta(\omega,T_2)}{\Theta(\omega,T_1)}\}}
{(1+|\tilde{\alpha}|^2)(1+\frac{\Theta(\omega,T_2)}
  {\Theta(\omega,T_1)})-2\Re\{\tilde{\alpha} + \tilde{\alpha}^*
  \frac{\Theta(\omega,T_2)}{\Theta(\omega,T_1)}\}}
\end{align*}
Here, the dimensionless, normalized polarizability
$\tilde{\alpha}=\alpha k_0^3 \kappa$ is introduced to capture the
dependence of radial spin $S_R$ on material properties (polarizability
$\alpha$), geometry (near-field coupling $\kappa$) and
wavelength/wavevector ($k_0$) in a concise manner. It follows that
when the normalized polarizability $\tilde{\alpha}=\pm i$, $S_R =
\pm\frac{\Theta(\omega,T_1)-\Theta(\omega,T_2)}{\Theta(\omega,T_1)
  +\Theta(\omega,T_2)}$. This gives high purity circular polarization
($S_R \rightarrow \pm 1$) when the ratio
$\Theta(\omega,T_2)/\Theta(\omega,T_1)$ is either very large or very
small. We illustrate with practical examples further below that this
dependence on both wavelength and temperature makes strong CP light
feasible even when temperatures $T_1,T_2$ are not very different. In
the following, not restricting ourselves to $\ev_z$ direction, we
explore the dependence of handedness on various design parameters.

\begin{figure}[t!]
  \centering \includegraphics[width=\linewidth]{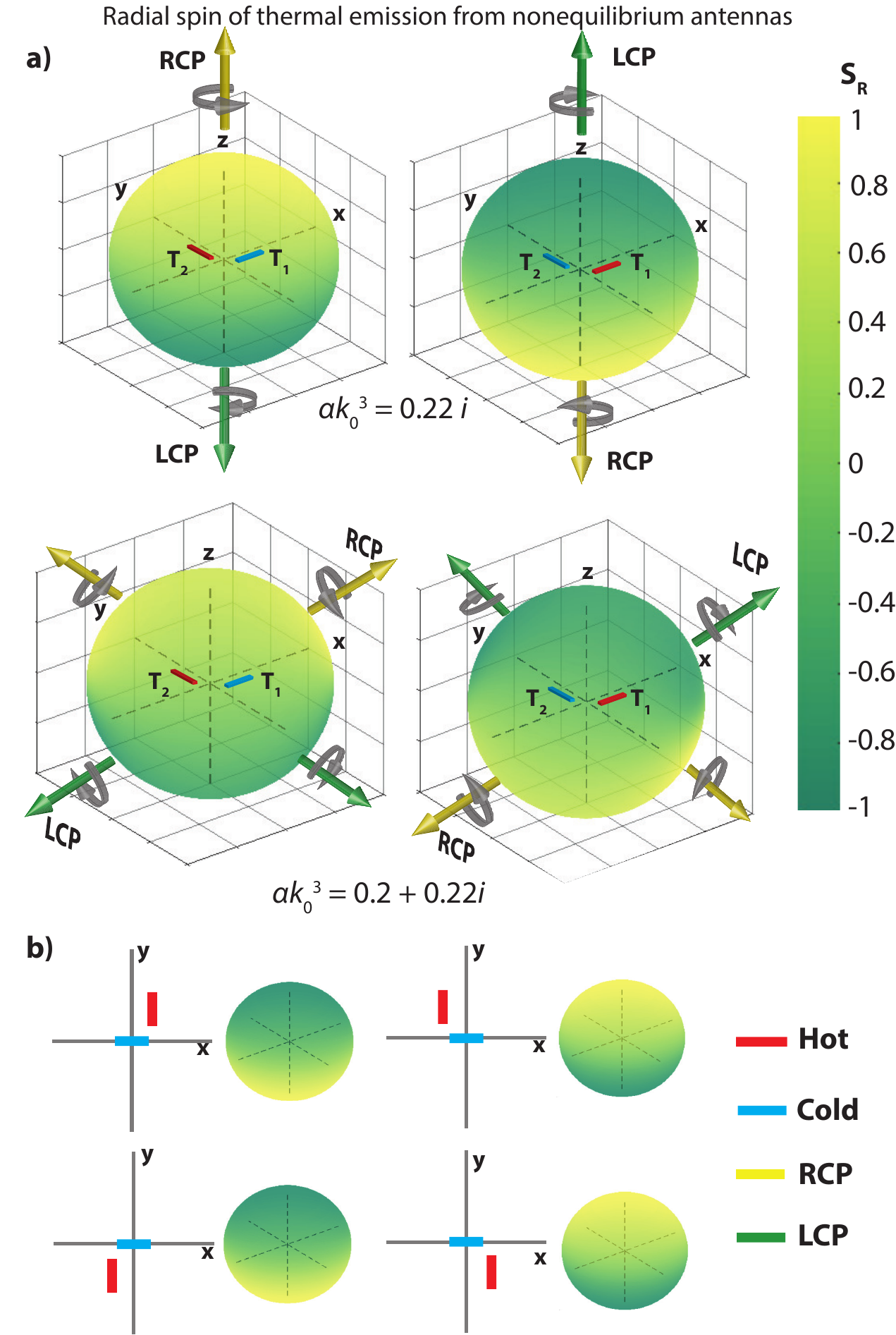}
  \caption{Radial spin of thermal emission $S_R$ from a dimer of two
    dipolar antennas is analyzed. Both antennas of dimensionless
    vacuum polarizability $\alpha k_0^3$ having temperatures $T_1,
    T_2$ and orientations $\ev_x, \ev_y$ respectively, are located at
    $\mathbf{x}_1=(0.1,-0.1,0)/k_0$ and
    $\mathbf{x}_2=(-0.1,0.1,0)/k_0$. (a) demonstrates the distribution
    of $S_R$ for two different $\alpha k_0^3$. Top two figures
    correspond to optimum design for maximum purity ($S_R=\pm 1$)
    along $\pm \ev_z$ directions while bottom two figures illustrate
    maximum purity along some intermediate directions for a different
    value of $\alpha k_0^3$. The handedness flipping RCP
    $\leftrightarrow$ LCP is observed upon interchanging temperatures
    $T_1 \leftrightarrow T_2$ or inverting emission direction $\ev_R
    \rightarrow -\ev_R$.  (b) illustrates the dependence of spin upon
    the relative spatial orientation of antennas. A mirror image
    configuration of a given dimer leads to opposite handedness (RCP
    $\leftrightarrow$ LCP flipping) in the same direction keeping all
    other parameters the same.}
  \label{neq}
\end{figure}

{\bf Dependence on antenna temperatures:} For small separation between
the antennas ($k_0d < 1$), it follows from Eq.\eqref{St} that $S_R
\propto \Im\{\langle\tilde{p}_1^*(\omega)\tilde{p}_2(\omega)
\rangle\}$. Under this condition, for any geometric configuration of
two nonequilibrium coupled antennas of equal polarizabilities
($\alpha_1=\alpha_2$), the thermal spin is flipped ($S_R \rightarrow
-S_R$) upon interchanging the temperatures of the antennas. While
there is no specific advantage of using antennas of unequal
polarizabilities ($\alpha_1\neq \alpha_2$), similar flipping of
handedness (sign of $S_R$) is observed but the change in the magnitude
of $S_R$ depends on the polarizabilities. As we show further below, it
is difficult to realize high purity ($|S_R|\sim 1$) CP radiation at
large separations due to decreased near-field interactions. But
similar tunability of handedness based on temperatures is observed at
large separations for weakly polarized light.

{\bf Dependence on emission direction:} We note that only one of the
normalized polarizability conditions $\alpha k_0^3 \kappa = \pm i$ is
true depending on the sign of near-field coupling $\kappa$ since the
condition $\Im\{\alpha\} > 0$ must hold true for real, passive (lossy)
dipoles~\cite{bohren2008absorption}. For the example configuration
with horizontal ($\ev_x$) and vertical ($\ev_y$) dipoles with relative
orientation $\ev_d=(-\ev_x+\ev_y)/\sqrt{2}$, the coupling $\kappa =
\frac{e^{i k_0 d}}{8\pi (k_0 d)^3}[(k_0d)^2-3+ 3ik_0d]$. Since
$|\Im\{\kappa\}/\Re\{\kappa\}| \ll 1$ and $\Re\{\kappa\}<0$ for
relevant separations $k_0d \leq 1$, it follows that $\tilde{\alpha} =
-i$ is the physically permissible optimum normalized
polarizability. Under this condition, radial thermal spin $S_R = +1$
(RCP) when $T_2 \gg T_1$ and $S_R = -1$ (LCP) when $T_2 \ll T_1$ for
emission direction $\ev_R =\ev_z$ (north pole). At $\ev_R=-\ev_z$
(south pole), the opposite handedness is observed under the same
conditions. Fig.\ref{neq}(a) demonstrates this dependence where the
two dipoles are assumed to be located at
$\mathbf{x}_1=(0.1,-0.1,0)\frac{1}{k_0}$ and
$\mathbf{x}_2=(-0.1,0.1,0)\frac{1}{k_0}$, having temperatures
$T_1,T_2$. For top two figures, the vacuum polarizability is $\alpha
k_0^3 =0.22i$ such that the optimum normalized polarizability
$\tilde{\alpha}=-i$ is realized. This results in RCP/LCP emission in
the northern/southern hemisphere when $T_2 \gg T_1$ and LCP/RCP
emission in northern/southern hemisphere when $T_2 \ll T_1$. This
flipping of thermal spin $S_R$ upon inverting the direction follows
from vectorial part of Eq.\eqref{St} where the radial spin $S_R
\rightarrow -S_R$ when $\ev_R \rightarrow -\ev_R$. As a consequence,
the total (integrated over all directions) angular momentum of emitted
radiation is zero as expected for a system lacking any intrinsic
source of angular momentum. In context of emission from magneto-optic
nanoparticles recently studied in Ref.~\cite{ott2018circular}, the
cylcotron motion of electrons in presence of applied magnetic field is
responsible for generating angular momentum intrinsically and
consequently, the total angular momentum radiated by that particle is
nonzero and lies along the direction of applied magnetic field (due to
its spherical, isotropic shape).

{\bf Dependence on polarizabilities:} The top two figures of
fig.\ref{neq}(a) illustrate the handedness distribution for
polarizability $\alpha k_0^3=0.22i$ which leads to the normalized
polarizability $\tilde{\alpha}=-i$, an optimum design for high purity
CP light along $\pm\ev_z$ direction. Bottom two figures of
Fig.\ref{neq}(a) illustrate the change in the handedness distribution
upon tuning the polarizability to $\alpha k_0^3=0.2+0.22i$. With this
latter configuration, the maximum purity of the radial spin occurs
along intermediate and multiple directions indicating that the
direction of maximum purity CP emission depends on the
polarizabilities and configuration of the antennas. This example shows
that the condition $\tilde{\alpha}=\pm i$ obtained above is not a
unique and limiting configuration for observing maximum purity CP
light from nonequilibrium antennas but other configurations can also
be employed.  For both these examples, despite complicated directional
dependence, northern/southern hemispheres contain the same overall
RCP/LCP emission when $T_2 \gg T_1$, with handedness flipped upon $T_2
\leftrightarrow T_1$. Here, we introduce the notation
$\leftrightarrow$ to denote the interchange of two quantities in a
concise manner. This allows us to summarize the above handedness
dependence as: RCP $\leftrightarrow$ LCP when $T_2 \leftrightarrow
T_1$ or $\ev_R \leftrightarrow -\ev_R$ keeping other parameters the
same.

{\bf Dependence on relative locations of antennas:} The overall
handedness for northern/southern hemispheres is determined by the
relative spatial configuration of dipoles $\ev_d$ which affects the
permissible optimum polarizability ($\tilde{\alpha}=+i$ or $-i$). In
particular, for the configuration discussed above, $\ev_d=(-\ev_x +
\ev_y)/\sqrt{2}$ leads to $\tilde{\alpha}=-i$ as the physically
permissible optimum design along $\pm \ev_z$. In an alternative
configuration with $\ev_d =(\ev_x + \ev_y)/\sqrt{2}$,
$\tilde{\alpha}=i$ is the physically permissible value, and this leads
to flipping of RCP $\leftrightarrow$ LCP in previous configurations
when all other parameters are kept the same. This is illustrated in
Fig.\ref{neq}(b) where L-shaped configuration in the $x-y$ plane with
hot vertical and cold horizontal antennas leads to RCP/LCP emission in
northern/southern hemisphere and opposite distribution is observed for
its mirror image (in $y-z$-plane) counterpart.

{\bf Dependence on orientations of antennas:} It further follows from
Eq.\eqref{St} that RCP $\leftrightarrow$ LCP is expected upon
interchanging the dipole orientations ($\ev_1 \leftrightarrow \ev_2$)
which has an important implication for any isotropic objects. For
isotropic objects, the fluctuating dipole moments along all three
directions are considered and the radial spin $S_R$ is calculated by
considering all dipole pairs. It turns out that the radial spin
vanishes for all directions due to spin cancellations from conjugate
pairs in the same plane. For instance, $S_R$ emitted by the pair of
dipoles $(\tilde{p}_{1x},\tilde{p}_{2y})$ is always opposite in
direction to that emitted by the pair
$(\tilde{p}_{1y},\tilde{p}_{2x})$ due to orientation
flipping. Moreover, they are equal in magnitude because $\langle
\tilde{p}_{1x}^* \tilde{p}_{2y}\rangle = \langle \tilde{p}_{1y}^*
\tilde{p}_{2x} \rangle$ which is true for equal polarizabilities along
both directions and provided that the interaction between dipoles is
reciprocal. From this, it follows that the anisotropic shape of
dipolar objects is necessary to produce CP light. Note that an
anisotropic nanoparticle by itself does not emit CP light due to lack
of correlations between orthogonal components of its fluctuating
dipole moments.

{\bf Dependence on distance between antennas:} While the dependence of
thermal spin on distance $d$ between the dipolar antennas is quite
complicated [Eq.\eqref{St}], the optimum polarizability condition
$\alpha k_0^3 \kappa = \pm i$ provides some useful insights. Since
$|\kappa|$ depends inversely on the distance $k_0 d$, it follows that
the polarizabilities ($\alpha k_0^3$) and the separation distance
($k_0 d$) required for the optimum design are also inversely
related. For small distances ($k_0 d \ll 1$), the phase factor
$e^{ik_0d(\ev_d\cdot\ev_R)} \sim 1$ in Eq.\eqref{St}. Under this
condition, one can realize maximum purity CP thermal emission with
small polarizabilities ($\alpha k_0^3 < 1$) using nonequilibrium
antennas discussed above. As we show below, the emission from the
antennas at equal temperatures in this regime is at best very weakly
polarized. For large separation distances ($k_0 d > 1$) corresponding
to negligible near-field interactions, the phase factor
$e^{ik_0d(\ev_d\cdot\ev_R)}$ matters and one can numerically optimize
the design. However, since $|\kappa| \propto \frac{1}{(k_0d)^3}$, the
required polarizabilities are very large ($\alpha k_0^3 \gg 1$) and
quite difficult to realize with real, practical systems.

{\bf Antennas at equal temperatures:} For small separation distances
$k_0d\ll 1$, the phase factor $e^{ik_0d(\ev_d\cdot\ev_R)} \sim 1$
simplifies Eq.\eqref{St}. It then follows that for two antennas of
polarizabilities $\alpha_1,\alpha_2$ having equal temperatures
($T_1=T_2$), radial spin $S_R \propto \Im\kappa
[\Im\alpha_1\Re\alpha_2 -\Im\alpha_2\Re\alpha_1]$. Thus, $S_R = 0$ not
only for equal vacuum polarizabilities ($\alpha_1=\alpha_2$) but also
when $\alpha_2/\alpha_1 \in \mathbb{R}$ where $\mathbb{R}$ stands for
real numbers. Even after overcoming the difficulty of achieving
$\alpha_2/\alpha_1 \notin \mathbb{R}$ with engineered design of
materials and geometrical shapes, it turns out that, such a dimer
produces very weakly polarized radiation. This occurs because of its
direct dependence on the dissipative (imaginary) part of near field
coupling ($S_R \propto \Im\{\kappa\}$) which is very small
($\Im\{\kappa\}\ll |\kappa|$) for relevant separations $k_0 d\lesssim
1$. For nonequilibrium antennas discussed above, the thermal spin
$S_R$ is non-negligible since it also depends on $\Re\{\kappa\}$
because of unequal temperatures. A dimer of antennas at equal
temperatures and having equal (or unequal) polarizabilities always
produces unpolarized (or very weakly polarized) thermal radiation.

We note that in order to produce strong CP thermal radiation with
antennas at equal temperatures, a many body system can be considered
where the antennas are anisotropic and arranged in a staggered manner
so that imaginary-valued correlations between orthogonal dipole
moments can be non-negligible through many-body interactions. A
metasurface of such anisotropic equilibrium antennas is already
explored in Ref.~\cite{shitrit2013spin} where a Kagome lattice of
silicon carbide nanorods is considered. In that work, the resulting
emission is predicted to be CP based on the spin-split dispersion of
underlying modes achieved with inversion asymmetric metasurface. While
the prediction is based on a concrete underlying mechanism, it can be
analyzed quantitatively by extending our fluctuational electrodynamic
analysis to a many-body system of such dipolar antennas.

{\bf Summary of fluctuational electrodynamic analysis:} In the
following, we point out the important aspects of CP thermal radiation
from nonequilibrium coupled antennas by summarizing the dependence of
radial spin $S_R$ upon a select few parameters analyzed above:
\begin{enumerate}
\item $S_R \leftrightarrow -S_R$ when $T_1 \leftrightarrow T_2$: The
  polarization state or handedness of emitted light can be
  reconfigured by interchanging the temperatures.
\item $S_R \rightarrow -S_R$ when $\ev_R \rightarrow -\ev_R$:
  Total angular momentum of emitted radiation is zero.
\item $S_R \leftrightarrow -S_R$ when $\ev_1 \leftrightarrow \ev_2$:
  Cancellation of spin of thermal emission from isotropic dipolar
  particles. Anisotropic shape is required for generation of CP light
\item The maximum purity of CP light from nonequilibrium antennas is
  $\text{max}\{S_R\}=\pm\frac{\Theta(\omega,T_1)-\Theta(\omega,T_2)}{
  \Theta(\omega,T_1) +\Theta(\omega,T_2)}$. Its magnitude depends only
  on the emission wavelength and the antenna temperatures while its
  direction depends on polarizabilities and geometric configuration of
  antennas.
\end{enumerate}

Thermal nonequilibrium enables strong CP thermal radiation from a
compact dimer of antennas that produces very weakly polarized emission
at equilibrium (equal temperatures). Strong CP thermal emission from
systems at thermal equilibrium is possible by using optimized chiral
absorber metasurfaces~\cite{wu2014spectrally} or many-body
configurations of
antennas~\cite{shitrit2013spin,yin2013interpreting}. However,
dynamical reconfigurability of circular polarization state with such
solid state designs is a major unsolved problem. Thermal
nonequilibrium enables the reconfigurability by simply interchanging
the temperatures. Furthermore, reconfigurability can be implemented at
arbitrary operating temperatures unlike other potential options of
reconfigurability such as phase change materials that are limited to
certain operating temperatures due to material specific transition
temperatures. Thermal nonequilibrium implemented in a dimer system of
antennas can thus achieve three important features of efficient CP
light sources namely, high purity, compactness and reconfigurability
of the polarization state. In the following, we explore the
experimental feasibility of this mechanism with suitable practical
examples.
  
\section{Practical Implementation}
\label{sec:example}

\begin{figure}[t!]
  \centering \includegraphics[width=\linewidth]{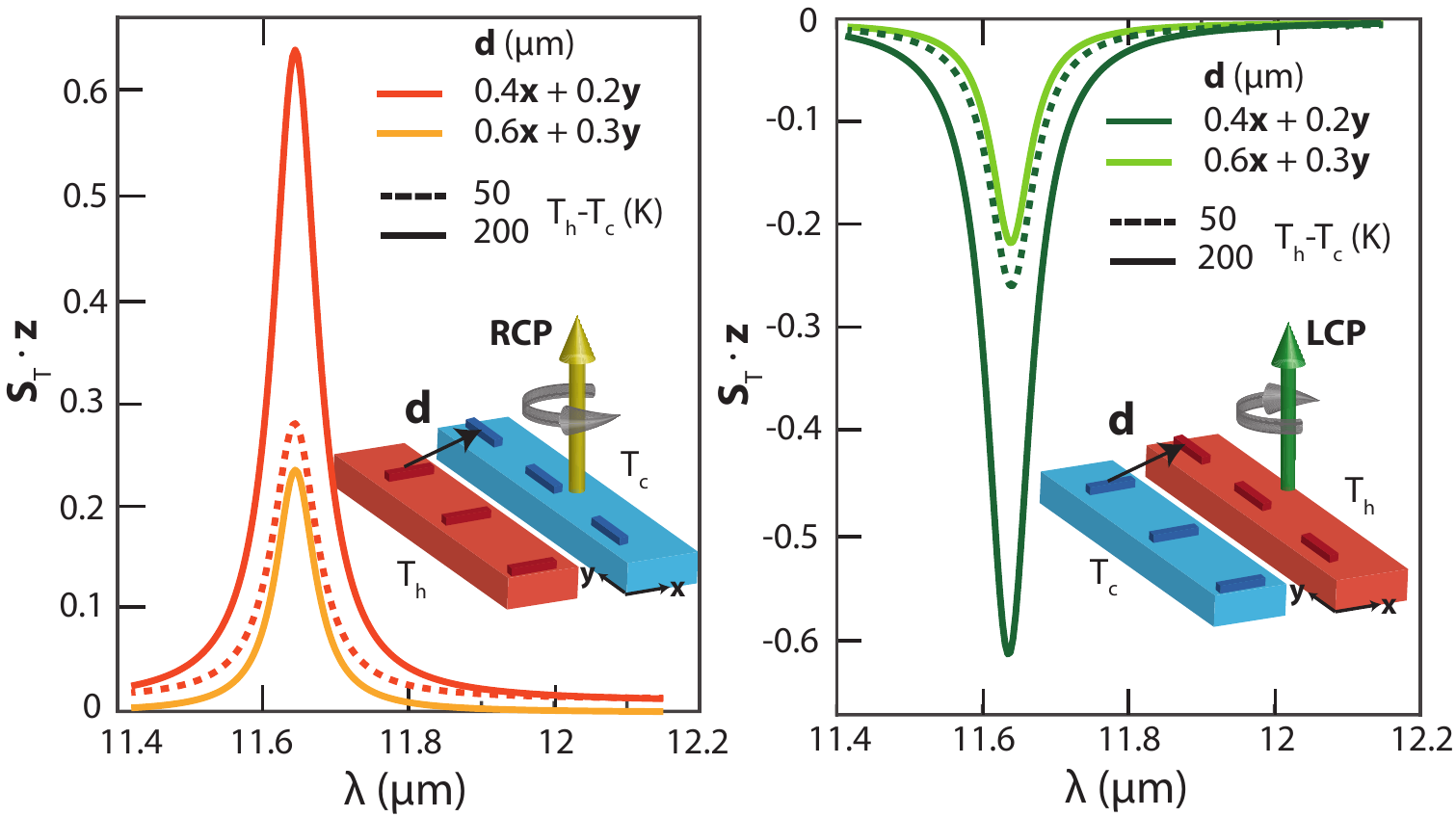}
  \caption{Two arrays of horizontally and vertically orientated
    Silicon Carbide nanoantennas spatially separated by $\mathbf{d}$
    and fabricated on top of two micro-heaters maintained at
    temperatures $T_c=300K$ (cold) and $T_h$ (hot). Length, breadth,
    height of each antenna are $0.4\mu$m, $0.1\mu$m,
    $0.1\mu$m. respectively. (a) When horizontal antennas are at
    hotter temperature than the vertical antennas, resulting emission
    is RCP. (b) The handedness of emission is switched to LCP when the
    temperatures are interchanged.}
  \label{fig2}
\end{figure}

As a potential experimental system to observe CP thermal emission from
nonequilibrium antenna dimers, we consider a system depicted in the
insets of Fig.\ref{fig2} comprising of two arrays of Silicon Carbide
(SiC) dipolar nanoantennas fabricated on top of suitable
micro-heaters. Silicon Carbide is considered a good material choice
for studying thermal emission due to large quality factor ($Q \sim
10^3$) phonon-polaritonic resonances which occur around
room-temperature thermal wavelengths ($\lambda \sim 10\mu$m). The
localized surface polaritonic modes supported by these anisotropic
antennas resonantly enhance the polarizabilities ($\alpha_1,\alpha_2$)
as well as the interactions between the antennas. We consider the
antennas having length, breadth and height of $0.4\mu$m, $0.1\mu$m and
$0.1\mu$m respectively. These can be approximated as prolate
ellipsoids~\cite{bohren2008absorption} having radii $R_a=200$nm (major
axis) and $R_b=50$nm (minor axis). For the realistic examples
considered here, the polarizability along the major axis is:
\begin{align}
\alpha(\omega) = \frac{4}{3}\pi R_a R_b^2
\frac{(\varepsilon(\omega)-1)}{1+N[\varepsilon(\omega)-1]}
\label{alpha}
\end{align}
where $N$ is the geometrical factor dependent on the eccentricity of
the ellipsoid $e_c=\sqrt{1-(R_b/R_a)^2}$ through the following
relation~\cite{bohren2008absorption}:
\begin{align*}
N = \frac{1-e_c^2}{e_c^2}\bigg(-1+\frac{1}{2
  e_c}\text{ln}\frac{1+e_c}{1-e_c}\bigg)
\end{align*}
$\varepsilon(\omega)$ is the permittivity of SiC taken from
reference~\cite{palik1998handbook} and is given below:
\begin{align}
\varepsilon(\omega) =
\varepsilon_{\infty}\bigg[\frac{\omega^2-\omega_{\text{LO}}^2+i\Gamma
    \omega}{\omega^2-\omega_{\text{TO}}^2+i\Gamma
    \omega}\bigg]
\label{restrahlen}
\end{align}
where $\varepsilon_{\infty}=6.7$,
$\omega_{\text{LO}}=1.825\times10^{14}$rad/s,
$\omega_{\text{TO}}=1.494\times 10^{14}$rad/s and
$\Gamma=8.966\times10^{11}$rad/s. The polarizability along minor axis
is given by \eqref{alpha} with the geometrical factor $N'=(1-N)/2$. It
is well-known that both polarizabilities are resonantly enhanced
(minimization of denominator) at two different frequencies. For the
purpose of illustration, we focus on CP thermal emission at
frequencies of dipolar resonant modes along major axes. At the
corresponding resonant wavelength, the polarizabilities along minor
axes are orders of magnitude weak and negligibly affect the dominant
thermal emission from polarizabilities along major axes.

The polarizability of these prolate ellipsoids is further influenced
by the presence of the substrate. We account for these changes within
the dipolar approximation using well-known Green's function
technique~\cite{novotny2012principles, amorim2017impact}. The modified
effective polarizability of the dipolar antennas is
$\alpha^{\text{eff}}(\omega)=\alpha(\omega)/[1-\alpha(\omega)k_0^3
  G(\omega)]$ where $k_0=\omega/c$. The expression
\begin{align}
G(\omega) = \frac{i}{8\pi}\int_0^{\infty} dk_p
\frac{k_p}{\sqrt{1-k_p^2}}[r^s-(1-k_p^2)r^p]e^{2i\sqrt{1-k_p^2}d_s}
\end{align}
is calculated numerically and it depends on the distance $d_s$ of the
antenna dipole from the surface of the metallic heater.  $r^s,r^p$
denote the usual Fresnel reflection coefficients for light incident
from vacuum onto the substrate. We assume the permittivity of the
heater to be $\varepsilon_{m}=-40+10i$ for calculation of Fresnel
reflection coefficients. We pick this arbitrary value to reflect the
fact that any highly reflective metallic surface such as tungsten,
doped semiconductor, metal nitrides can be used in its place. The
dependence of the results upon changes in the large negative substrate
permittivity is negligible. We note that the dipolar analysis is not
accurate for large field gradients and higher order multipoles
contributing when the antennas are separated by small
surface-to-surface (antenna surface) separations $d_s$ such that
$d_s/R_a < 1$. We avoid this regime for illustration purpose.


Figure~\ref{fig2} describes the thermal emission spin ($S_T$) along
the normal direction ($\ev_z$) from the dimer pairs. It shows its
dependence upon the separation between the antennas $\mathbf{d}$ and
their temperatures, $T_c=300K$ (cold) and $T_h=T_c + \Delta T$
(hot). We assume that the dimer pairs are separated by distance much
larger than $\mathbf{d}$ and use the arrays to motivate an actual
experiment. The magnitude of $S_T \cdot \ev_z$ for two different
separations is shown by light and dark orange (a) and green (b) curves
and that for two different $\Delta T$ is shown by solid and dashed
lines. As depicted in the inset of Fig.\ref{fig2}(a), we first
consider a configuration where horizontal dipoles (in $x-y$ plane) are
heated to high temperature $T_h$ while the vertical dipoles are
maintained at a lower temperature $T_c$. This results in the emission
of RCP light along the normal $\ev_z$ direction. When the temperatures
of the antennas are flipped leading to a configuration depicted in the
inset of Fig.\ref{fig2}(b), the resulting emission is LCP keeping all
other parameters the same. Since microheater temperatures can be
tuned, this enables reconfigurability of the polarization state of the
emitted radiation.

As described earlier, for given temperatures $T_h$ and $T_c$, the
maximum purity that can be reached upon design optimization is
$\frac{\Theta(\omega,T_h)-\Theta(\omega,T_c)}{\Theta(\omega,T_h)+
  \Theta(\omega,T_c)}$ and depends on both wavelength and
temperatures. At mid-infrared wavelength of $11.6\mu$m, the proposed
device emits strong circularly polarized light with thermal spin $S_T
\sim 0.65$ for $\Delta T=200$K and $S_T \sim 0.3$ for $\Delta
T=50$K. Note that the SiC antennas also exchange near-field radiative
heat flux of the order of $\lesssim 1$nW for these configurations (not
shown). This near-field heat flux affects the temperatures of
antennas. This effect can be monitored in a real experiment and can be
analyzed using a separate thermal model. Here, we use reasonable
values of steady state temperatures of the antennas for the
calculations. Overall, this analysis of the thermal spin of emission
from SiC antenna dimers indicates feasibility of observing reasonably
strong CP light under temperature difference of $\Delta T \sim 50$K
across separation of $d \gtrsim 0.5\mu$m between the antennas.

\begin{figure}[t!]
  \centering \includegraphics[width=\linewidth]{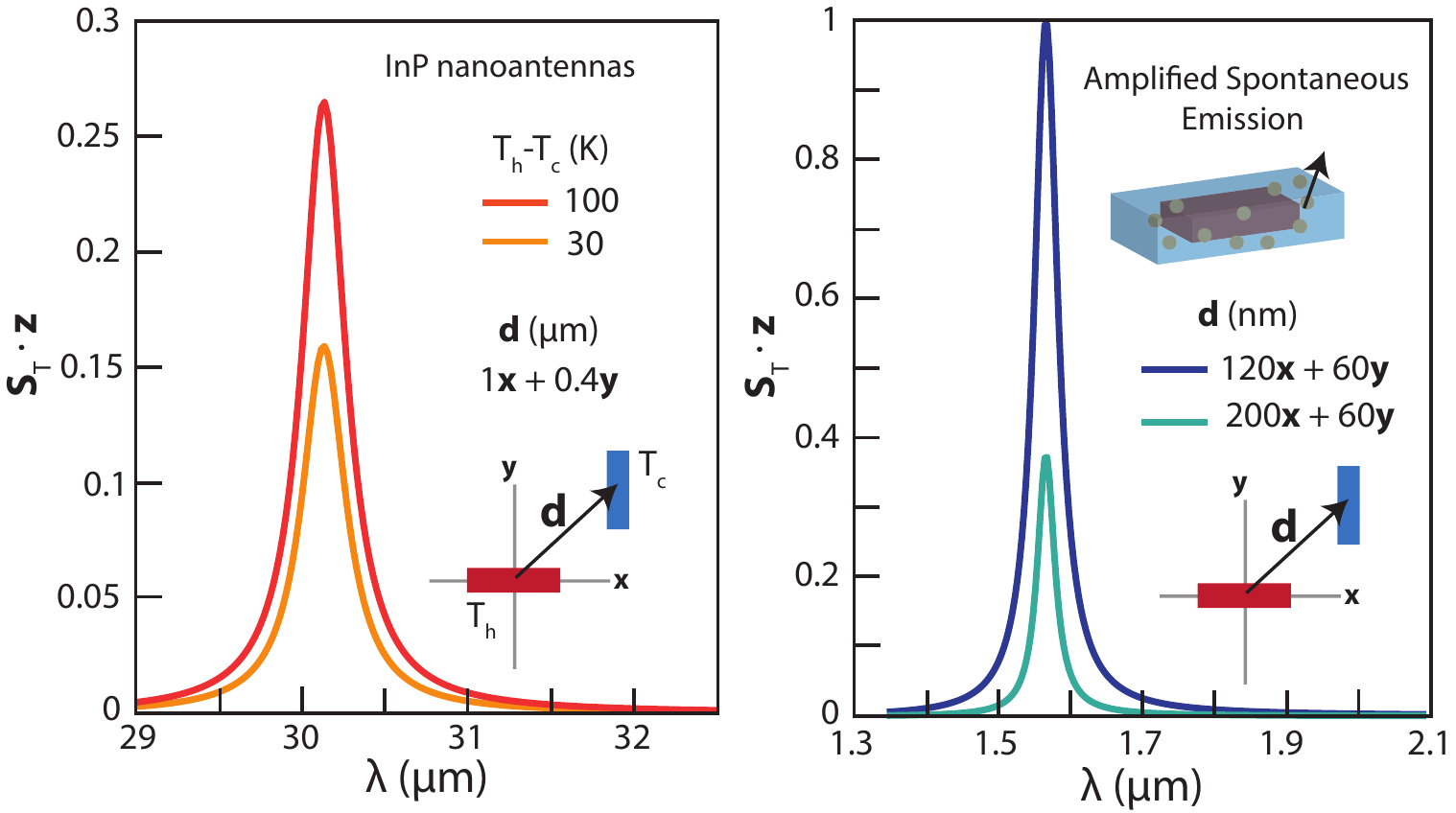}
  \caption{(a) Thermal emission spin ($S_T$) for a dimer of InP
    antennas illustrates feasibility of observing mid-IR CP light with
    slightly larger gap size for similar $\Delta T=T_h-T_c$, compared
    to SiC antennas analyzed in figure~\ref{fig2}. This figure shows
    that CP light generation from nonequilibrium antennas is not
    limited to one particular material choice. Figure (b) demonstrates
    a conceptual example of a dimer system of antennas where one of
    the antennas contains gain medium i.e. a plasmonic antenna
    enclosed in a dye-doped shell. Well below the lasing threshold,
    the incoherent amplified spontaneous emission (ASE) from the gain
    antenna coupled with another nearby antenna can lead to high-
    purity CP light at suitable separations. This example shows that
    the proposed mechanism is not limited to ``thermal'' nonequilibrium
    but can be potentially implemented in other forms such as
    population inversion of gain atoms.}
  \label{fig3}
\end{figure}

One can also consider other Restrahlen
materials~\cite{caldwell2015low} such as Indium Phosphide (InP),
Gallium Arsenide (GaAs), Indium Antimonide (InSb) etcetera which
support localized polaritons at longer wavelengths and can allow
generation of CP light for larger separations. As an example,
figure~\ref{fig3}(a) demonstrates CP-light generation from a dimer of
coupled nonequilibrium InP nanoantennas. The permittivity of InP is
obtained from Ref.~\cite{palik1998handbook} and is described by the
Lorentz oscillator model given in Eq.\eqref{restrahlen} with following
parameteres. $\varepsilon_{\infty}=9.61$,
$\omega_{\text{LO}}=6.498\times 10^{13}$rad/s,
$\omega_{\text{TO}}=5.720\times 10^{13}$rad/s and
$\Gamma=6.495\times10^{11}$rad/s. Length, breadth and height of each
antenna are assumed to be $0.8\mu$m, $0.4\mu$m,
$0.4\mu$m. respectively. Figure~\ref{fig3}(a) shows the spectral
thermal spin of emitted radiation when the antennas are separated by a
fixed separation of $d \approx 1.1\mu$m with temperatures $T_c=100$K
and $T_h$ (hot). As shown, the thermal emission spin $S_R \sim 0.15$
is obtained for a temperature difference of $\Delta T=30$K, indicating
the feasibility of observing CP light with smaller $\Delta T$ and
larger $d$ compared to SiC antennas. This example shows that other
materials, operating temperatures and dimer configurations can be
explored to optimize emission of CP light from nonequilibrium
antennas.

While we focused so far on thermal nonequilibrium, other forms of
nonequilibrium can be potentially implemented using the same dimer
configuration. One example is amplified spontaneous emission (ASE)
from one of the antennas containing active gain medium where
nonequilibrium is in the form of population inversion of gain
atoms. The incoherent ASE noise below loss compensation (lasing
threshold) is described using fluctuational electrodynamic
theory~\cite{matloob1997electromagnetic,henry1996quantum}. It arises
from the fluctuating dipole moments associated with the atomic
transitions of gain atoms. These fluctuating sources have an effective
temperature dependent on the population inversion and their
correlations are described using specific fluctuation-dissipation
relations~\cite{matloob1997electromagnetic,khandekar2016giant}. As
depicted in the inset of fig.\ref{fig3}(b), we consider a metallic
(e.g. gold, silver) antenna enclosed in a dye-doped shell which acts
as an amplifying medium. Such intricate systems have been considered
experimentally in context of plasmonic
nanolasers~\cite{noginov2009demonstration,stamplecoskie2014self}. While
large pump requirements for plasmonic lasing in these geometries make
the experimental realization of lasing
challenging~\cite{premaratne2017theory}, here we operate with lower
gain values (well below lasing threshold) and consider its use for
circularly polarized ASE from the dimer.

For this conceptual example, we consider the gain nanoantenna to have
an effective permittivity:
\begin{align}
  \varepsilon(\omega) =
  \underbrace{1-\frac{\omega_p^2}{\omega_p^2+i\Gamma\omega_p}}_{\varepsilon_r}
  + \underbrace{\frac{D_0\gamma_{\perp}}{(\omega-\omega_{21})+
      i\gamma_{\perp}}}_{\varepsilon_g}
\end{align}
consisting of Drude part $\varepsilon_r$ and a gain medium part
$\varepsilon_g$. The transition frequency of gain atoms is
$\omega_{21}$ while the strength of (pump-tunable) population
inversion is characterized by $D_0$. We choose the parameters
$\omega_p=4.38\times 10^{15}$rad/s, $\Gamma=\omega_p/200$,
$\omega_{21}=1.2\times 10^{15}$rad/s,
$\gamma_{\perp}=\omega_{21}/100$.  The gain is chosen to be $D_0=0.15$
such that $\Im\{\varepsilon\}>0$ at all wavelengths ensuring ASE
regime well below loss-compensation. We assume the other nanoantenna
to be characterized by the permittivity $\varepsilon_r$. Length,
breadth and height of each antenna are $100$nm, $25$nm, $25$nm
respectively. The polarizabilities $\alpha_{1,2}(\omega)$ are
calculated using Eq.\eqref{alpha}. The fluctuating polarization
associated with ASE from the gain antenna is described by the
fluctuation dissipation relation~\cite{khandekar2016giant}:
\begin{align}
  \langle p_{1g}^*(\omega)p_{1g}(\omega')\rangle =
  \frac{2\eps_0}{\omega}
  \text{Im}\{\alpha_1(\omega)\}\frac{\Im\varepsilon_g(\omega)}
       {\Im\varepsilon(\omega)} \frac{-n_2\hbar\omega_{21}}{n_2-n_1}
       \delta_{\omega,\omega'}
\end{align}
where $n_2,n_1$ represent the populations in excited and ground states
of gain atoms. We assume $n_2=0.95n$, $n_1=n-n_2$ with $n$ as the
total number of gain atoms. Since ASE is much larger than the thermal
radiation from the antennas, it suffices to calculate the spectral
thermal spin of ASE given by equation \eqref{St} and calculated using
above correlations. Figure~\ref{fig3}(b) demonstrates the spectral
thermal spin for two different separations $\mathbf{d}$ between the
antennas. For the above parameters, the highest purity CP light
($S_R=1$) is obtained at a wavelength $\lambda = 1.55\mu$m, for
$d=134$nm as shown by the dark blue curve. This conceptual example
shows that the proposed mechanism for generating CP light using
nonequilibrium antennas is not limited to systems under thermal
nonequilibrium. Apart from the use of active gain medium, one can also
consider use of biased semiconductors where the heat bath of
underlying fluctuating currents has an effective chemical potential or
temperature that can be tuned by changing the voltage
bias~\cite{henry1996quantum,chen2015heat}. These other forms of
nonequilibrium requiring more
experimental~\cite{noginov2009demonstration,stamplecoskie2014self} and
theoretical details can be explored in separate future works.

{\bf Summary of potential experimental platforms:} For mid-IR CP
thermal emission based on thermal nonequilibrium, one important
technical difficulty is that of maintaining large temperature
difference across small gaps for high-purity CP light. However, we are
confident that the proposed mechanism of CP-light generation can be
experimentally implemented in the near future. In particular, there
has been a significant experimental progress in the area of near-field
radiative heat transfer which has facilitated exploration of thermal
nonequilibrium at nanoscale and has also verified the validity of
fluctuational electrodynamic
theory~\cite{song2015near,kim2015radiative}. In particular, much
larger temperature differences ($\Delta T \gtrsim 100K$) across small
gaps $d \lesssim 1\mu$m. have been experimentally probed in suitable
geometries~\cite{bernardi2016radiative,ghashami2018precision}. Our
analysis of SiC and InP antennas above provides an estimate that
reasonably strong CP light ($S_T \sim 0.2$) can be observed with such
dimers held at temperature difference as small as $\Delta T \sim 30K$
and separated by a gap size of $d \sim 1\mu$m. Such an experiment will
be fundamentally important as it will reveal for the first time the
non-intuitive connection between thermal nonequilibrium and angular
momentum of light.

In the long run, CP-light emission at near-IR wavelengths and other
practical applications such as CP light emitting LEDs can be pursued
by exploring other nonequilibrium systems considered above. In
comparison to other CP light sources based on structural chirality or
polarization conversion, reconfigurability is an important distinct
advantage of the proposed mechanism. Related to the tunability of
device temperature in context of CP thermal emission, recent
works~\cite{sakat2018enhancing,mori2014electrically,kim2018ultrafast}
have explored dynamical modulation of thermal emission in
two-dimensional opto-electronic platforms where ultra-fast switching
rates $\gtrsim 100$MHz have been realized. We therefore envision that
in the long run, the proposed mechanism can be combined with such
newly emerging concepts to build a compact, high purity and
dynamically reconfigurable source of CP light.

\section{Conclusion} \label{sec:conclusion}

We demonstrate a mechanism of CP thermal radiation based on near-field
coupling between nonequilibrium antennas. It is unrelated to geometric
and material chirality or use of any polarization conversion
device. We show that a simple dimer of coupled nonequilibrium antennas
can facilitate a great degree of control over the polarization of
emitted radiation upon tuning the temperatures, emission direction,
relative orientations and positions of antennas. Through a rigorous
fluctuational electrodynamic analysis of the dependence on these
parameters, we reveal the fundamental origin and describe the general
design guidelines for generating CP light from fluctuating thermal
sources. Our analysis revealed that the imaginary valued correlations
between orthogonal components of fluctuating sources are necessary for
emission of CP light. In the context of thermal emission from coupled
dipolar thermal sources, anisotropic shape is necessary to emit CP
thermal radiation. While the computational design with such dipolar
bodies is simple and convenient, one can also go beyond this regime
with advanced computational tools and inquire about the circular
polarization described by Eq.\eqref{spin} of thermal emission from
arbitrary, non-intuitive geometries~\cite{reid2015efficient}.

We further explore the experimental feasibility of generating CP light
from nonequilibrium antennas with realistic examples of SiC and InP
antennas. A reasonably strong CP light from nonequilibrium antennas
can be detected in the near future using these example systems. To
show that the proposed mechanism is not limited to ``thermal''
nonequilibrium, we further provide a conceptual example of plasmonic
nanolaser system that can emit circularly polarized amplified
spontaneous emission by operating well below lasing
threshold. Consideration of such nontrivial approaches is important
for practical applications such as CP-light-emitting LEDs.

The underlying approach of analyzing spin angular momentum property of
thermal radiation remains largely unexplored in the field of thermal
radiation~\cite{ott2018circular}. This approach opens the door to
numerous future studies of angular-momentum-related radiative heat
transport phenomena. For instance, one immediate extension of the
current work can be consideration of many-body configurations of
anisotropic dipolar antennas for shaping angular-momentum properties
of light, using thermal nonequilibrium without applying magnetic
field. Our analysis also shows a spatial distribution of angular
momentum radiated by nonequilibrium antennas. This implies that
nontrivial torques on nanoscale bodies can be expected in the vicinity
of these antennas. Interestingly, this is already probed by other
recent work~\cite{reid2017photon} which demonstrates thermal
nonequilibrium enabled torques with temperature dependent sign and
magnitude. This surprising fundamental connection between thermal
nonequilibrium and angular momentum of light suggests new
possibilities in the context of not only nonequilibrium but also
nonisothermal bodies. From the perspective of nanoscale thermometry,
detection and sensing applications, one question that arises is how
angular momentum of emitted radiation is influenced by the temperature
gradients in nonisothermal bodies which requires an answer necessarily
within fluctuational electrodynamic theory and yet remains
unsolved. We leave these inquiries aside for future work.

\section{Acknowledgments}

We would like to thank Ryan Starko-Bowes and Aman Satija for helpful
discussions. This work was supported by the U.S. Department of Energy,
Office of Basic Energy Science under award number DE-SC0017717 and the
Lillian Gilbreth Postdoctoral Fellowship program at Purdue University
(C.K.).

\bibliography{photon}

\end{document}